\documentstyle[12pt]{article}
\newcommand\be{\begin{equation}}
\newcommand\ee{\end{equation}}
\newcommand\bea{\begin{eqnarray}}
\newcommand\eea{\end{eqnarray}}
\newcommand\ket[1]{|#1\rangle}

\newcommand{\fatalpha}{{\bf \alpha \kern -0.44em \alpha}}
\newcommand{\fatsigma}{{\bf \sigma \kern -0.54em \sigma}}
\newcommand{\tpchi}{{\bf \chi \kern -0.35em \chi}}
\newcommand{\llambda}{{\bf \lambda \kern -0.45em \lambda}}

\bibliography{plain}
\title{\bf Perfect state transfer over interacting boson networks associated with group schemes}\vspace{20mm}
\author{ M. A. Jafarizadeh$^{a,b,c}$
 \thanks{E-mail:jafarizadeh@tabrizu.ac.ir},
 R. Sufiani$^{a,b}$
 \thanks{E-mail:sofiani@tabrizu.ac.ir}, M. Azimi$^{a}$ and F. Eghbali Fam$^{a}$
 \\ $^a${\small Department of Theoretical Physics and Astrophysics,
University of Tabriz, Tabriz 51664, Iran.} \\ $^b${\small
Institute for Studies in Theoretical Physics and Mathematics,
Tehran 19395-1795, Iran.} \\ $^c${\small Research Institute for
Fundamental Sciences, Tabriz 51664, Iran.}} \pagebreak

\vspace{20mm}
\begin{document}
\maketitle \vspace{15mm}
\newpage
\begin{abstract}
It is shown how to perfectly transfer an arbitrary qudit state in
interacting boson networks. By defining a family of Hamiltonians
related to Bose-Hubbard model, we describe a possible method for
state transfer through bosonic atoms trapped in these networks
with different kinds of coupling strengths between them.
Particularly, by taking the underlying networks of so called group
schemes as interacting boson networks, we show how choose suitable
coupling strengths between the nodes, in order that an arbitrary
qudit state be transferred from one node to its antipode,
perfectly. In fact, by employing the group theory properties of
these networks, an explicit formula for suitable coupling
strengths has been given in order that perfect state transfer
(PST) be achieved. Finally, as examples, PST on the underlying
networks associated with cyclic group $C_{2m}$, dihedral group
$D_{2n}$, Clifford group $CL(n)$, and the groups $U_{6n}$ and
$V_{8n}$ has been considered in details.


{\bf PACs Index: 01.55.+b, 02.10.Yn }
\end{abstract}

\vspace{70mm}
\newpage
\section{Introduction}
Transfer a quantum state from one site to another one (quantum
communication between two parts of a physical unit, e.g., a qubit)
is a crucial ingredient for many quantum information processing
protocols (QIP)\cite{1}. However, in many other tasks, e.g.,
solid-state-based quantum computation, quantum state transmission
is not a trivial task. Completing such a task is often needed in
QIP, e.g., the quantum information exchange between two separate
processors. Therefore, it is very important to find physical
systems that provide robust quantum state transmission lines
linking different QIP processors. There are various physical
systems that can serve as quantum channels, one of them being a
quantum spin system. In recent years, many results have been
obtained on qubit-state transfer through spin chains with various
types of neighbor couplings. The original idea of quantum state
transfer through a system of interacting spin $1/2$ was introduced
by Bose \cite{Bose,5}. Afterward, Christandl et al. \cite{8,9''}
and independently Nikolopoulos et al. \cite{nik} found that
perfect state transfer (PST) is possible \cite{bu2} in spin-1/2
networks without any additional actions from senders and
receivers, including not requiring switching on and off qubit
couplings. Similarly, in Ref. \cite{6}, near perfect state
transfer was achieved for uniform couplings provided a spatially
varying magnetic field was introduced. After the work of Bose
\cite{Bose}, in which the potentialities of the so-called spin
chains have been shown, several strategies were proposed to
increase the transmission fidelity \cite{Os} and even to achieve,
under appropriate conditions, perfect state transfer
\cite{8,9'',Bu,Bu1,yung,yung1, Facer}. Recently, A. Bernasconi, et
al. \cite{godsil} have studied perfect state transfer (PST)
between two particles in quantum networks modeled by a large class
of cubelike graphs. Since quantum networks (and communication
networks in general) are naturally associated to undirected
graphs, one can use the relation between graph-theoretic
properties and properties that allow PST \cite{godsil1}.

In Ref. \cite{PST}, the so called distance-regular graphs have
been considered as spin networks (in the sense that with each
vertex of a distance-regular graph a qubit or a spin $1/2$
particle was associated) and PST over them has been investigated.
In the paper \cite{psd}, optimal state transfer (ST) of a
$d$-level quantum state (qudit) over pseudo distance regular
networks was investigated, where it was shown that only for pseudo
distance regular networks with some certain symmetry (mirror
symmetry) in the corresponding intersection numbers, PST between
antipodes of the networks can be achieved. Also, in the recent
paper \cite{pstass}, PST of a qudit state in a system of $N$ spin
$j$ particles located at the nodes of an underlying network of a
so-called group association scheme \cite{Ass.sch.} has been
studied. In paper Ref. \cite{wu}, PST of any qudit state through
bosonic lattices has been investigated. They have considered a
model which can be implemented using the Bose-Hubbard model and
proposed a protocol to perfectly transfer an unknown $n$-variable
function from a processor at one end of a boson chain to another
processor at the other.

In this work, we consider the underlying networks of group
schemes, as interacting boson networks and impose a more general
Bose-Hubbard Hamiltonian to the considered networks. More clearly,
for a given vertex set (each vertex is associated with an element
of a finite group $G$) we define different adjacency matrices
according to different kinds of coupling strengths between the
vertices (nodes). Then, we investigate perfect state transfer
(PST) on the nodes of these networks and show that by choosing
suitable coupling strengths between the nodes, an arbitrary qudit
state can be transferred from one node to its antipode, perfectly.
In fact, by employing the group theory properties of these
networks, we give an explicit formula for suitable coupling
strengths in terms of irreducible characters of the corresponding
groups. As examples, we consider PST on the underlying networks
associated with cyclic group $C_{2m}$, dihedral group $D_{2n}$,
Clifford group $CL(n)$, and the groups $U_{6n}$ and $V_{8n}$, in
details.

The organization of the paper is as follows: In section 2, PST
over antipodes of interacting boson networks is investigated,
where a method for finding suitable coupling constants in
particular Bose-Hubbard Hamiltonians so that PST be possible, is
given. Section $3$ is devoted to PST on special networks called
the underlying networks of group schemes, where a formula for
suitable coupling strengths between nodes is given in order that
PST of a qudit state from an arbitrary node to its antipode be
achieved. In section $4$, some important examples of underlying
networks of group schemes are considered and PST over them is
investigated in details. The paper is ended with a brief
conclusion.
\section{Perfect state transfer in interacting boson networks}
Consider the Bose-Hubbard model in which dynamics of bosons, in a
system with $N$ sites, governed by a linearly coupled bosonic
Hamiltonian as
\begin{equation}\label{eqH0}
H=\sum_{k=1}^{N-1}J_{k}({b^{\dag}}_kb_{k+1}+{b^{\dag}}_{k+1}b_{k})+\sum_{k=1}^{N}{\epsilon}_kn_k
\end{equation}
where, $n_k={b_k}^{\dag}b_k$ is the number operator for the bosons
located at the $k$th site, ${b_k}^{\dag}$ ($b_k$) is the bosonic
creation (annihilation) operator. For simplicity, $\hbar=1$ has
considered.

We generalize the above model to more general finite graphs not
only for a finite path, in a way that the nodes of graph have
different kinds of coupling strengths between themselves. Suppose
$\Gamma$ is a connected graph. For each coupling strength $J_k$,
$k=0,1, \ldots, d$, we can form a graph $\Gamma_k$ in which
vertices are adjacent if their coupling in $\Gamma$ equals $k$.
Let $A_k$ be the adjacency matrix of $\Gamma_k$. For instance,
$A_1$ is the adjacency matrix $A$ of $\Gamma$. Also, let $A_0=I$,
the identity matrix. This gives us $d + 1$ matrices $A_0, A_1,
..., A_d$, called the adjacency matrices of $\Gamma$. Their sum is
the matrix $J$ in which every entry is $1$. In the other words, we
assume that the dynamics of bosons, in a system with $N$ sites
(associated with the nodes of a finite group), is governed by the
following Bose-Hubbard Hamiltonian
\begin{equation}\label{eqH}
H=\sum_{i,j=1}^{N}\sum_{k=0}^{d}J_{k}(A_k)_{ij}b^{\dag}_ib_j.
\end{equation}

Now, we assume that the adjacency matrices $A_k$ for
$k=0,1,\ldots, d$ commute with each other such that, one can
diagonalize them by a unitary matrix $U$, simultaneously. That is,
we have
$$UA_kU^{\dag}=D_k,$$
where
$D_k=diag(\lambda^{(k)}_1,\lambda^{(k)}_2,\ldots,\lambda^{(k)}_{N})$
is a diagonal matrix with eigenvalues of $A_k$ as its diagonal
entries. Therefore, the Hamiltonian (\ref{eqH}) can be rewritten
as
\begin{equation}\label{eqH1}
H=\sum_{i,j}\sum_kJ_{k}(U^{\dag}D_kU)_{i,j}b^{\dag}_ib_j=\sum_{k=0}^{d}\sum_{l=1}^{N}J_{k}\lambda^{(k)}_l\sum_{i}U^{\dag}_{il}b^{\dag}_i\sum_{j}U_{lj}b_j
\end{equation}
By considering the Bogoliubov transformations
\begin{equation}\label{eqH2}
{\tilde{b}}_l=\sum_{j}U_{lj}b_j, \;\;\
b_l=\sum_{j}U^{*}_{jl}{\tilde{b}}_j , \;\;\ l=1,2,\ldots, N
\end{equation}
and defining ${\tilde{n}}_l={\tilde{b}}^{\dag}_l{\tilde{b}}_l$,
the Hamiltonian (\ref{eqH1}) is written as
\begin{equation}\label{eqH3}
H=\sum_{l=1}^{N}{\tilde{J}}_{l}{\tilde{n}}_l,
\end{equation}
where,
\begin{equation}\label{eqH4}
{\tilde{J}}_{l}=\sum_{k=0}^{d}J_k\lambda^{(k)}_l.
\end{equation}
Now, assume that we are able to prepare the initial state
$\ket{\phi_0}=\alpha\ket{{\mathbf{0}}}+\beta
b^{\dag}_1\ket{{\mathbf{0}}}$ (with $|\alpha|^2+|\beta|^2=1$ and
$\ket{{\mathbf{0}}}:=\ket{00...00}$ as the vacuum state). That is,
we consider that only the first site is occupied. Then, the
network couplings are switched on and the system is allowed to
evolve under $U(t)=e^{-iHt}$ for a fixed time interval, say $t_0$.
The final state becomes \be \label{eqH5}\ket{\phi(t_0)}
=e^{-iHt_0}\ket{\phi_0}=
\alpha\ket{{\mathbf{0}}}+\beta\sum_{j=1}^Nf_{j1}(t_0)b^{\dag}_j\ket{{\mathbf{0}}}
\ee where, $f_{j1}(t_0):=\langle
{\mathbf{0}}|b_je^{-it_0\sum_{l=1}^{N}{\tilde{J}}_{l}{\tilde{n}}_l}b^{\dag}_1\ket{{\mathbf{0}}}$.
We consider perfect state transfer, i.e., we want to transfer
informations coded in the first site (starting site) to the site
$m$ (target site), perfectly. This means that, we impose the
condition \be\label{eq3} |f_{m1}(t_0)|=1\;\;\ \mbox{for}\;\
\mbox{some}\;\ 0<t_0<\infty\ee
 which can be interpreted as the signature of
perfect communication (or PST) between sites $1$ and $m$ in time
$t_0$. In order to achieve this condition, we use the identity
$b^{\dag}_1=\sum_lU_{l1}{\tilde{b}}^{\dag}_l=\sum_l{\tilde{b}}^{\dag}_l$
to rewrite the evolved state (\ref{eqH5}) as \be
\label{eqH5'}\ket{\phi(t_0)} =
\alpha\ket{{\mathbf{0}}}+\beta\sum_{l=1}^{N}e^{-it_0{\tilde{J}}_{l}}{\tilde{b}}^{\dag}_l\ket{{\mathbf{0}}}=\alpha\ket{{\mathbf{0}}}+\beta\sum_{j,l=1}^{N}e^{-it_0{\tilde{J}}_{l}}U^{*}_{lj}{b}^{\dag}_j\ket{{\mathbf{0}}}
\ee where, we have used the Eq.(\ref{eqH2}). By defining column
matrix $\mathbf{\tilde{J}}$ as
\begin{equation}\label{J}{\mathbf{\tilde{J}}}=\left(\begin{array}{c}
                      e^{-it_0{\tilde{J}}_{1}} \\
                         e^{-it_0{\tilde{J}}_{2}} \\
                         \vdots \\
                         e^{-it_0{\tilde{J}}_{N}} \\
                       \end{array}\right),
                       \end{equation}
                       the final state $\ket{\phi(t_0)}$ can be
                       written as
\be\label{eqH5''}\ket{\phi(t_0)} =
\alpha\ket{{\mathbf{0}}}+\beta\sum_{j}(\sum_{l}U^{\dag}_{jl}{\mathbf{\tilde{J}}}_l){b}^{\dag}_j\ket{{\mathbf{0}}}=\alpha\ket{{\mathbf{0}}}+\beta\sum_{j=1}^{N}{({U^{\dag}{\mathbf{\tilde{J}}}})}_j{b}^{\dag}_j\ket{{\mathbf{0}}}
\ee By comparing the above equation with (\ref{eqH5}) we see that
\be\label{f}
f_{j1}(t_0)={({U^{\dag}{\mathbf{\tilde{J}}}})}_j=\sum_{l=1}^N
U^{\dag}_{jl}e^{-it_0{\tilde{J}}_l}.\ee
 Now, in order to achieve PST to the $m$-th site, it is
sufficient to have \be
\label{eqH6}{({U^{\dag}{\mathbf{\tilde{J}}}})}_i=e^{i\theta}\delta_{im}.\ee

In order to obtain some informations about the matrix $U$, we
project the Hamiltonian $H$ to the single particle subspace, in
which the sites are empty or occupied by only one boson. Then, by
defining the kets $\ket{i_1,i_2,...,i_N}$ with $i_1,...,i_N\in
\{0,1\}$ as an orthonormal basis for Hilbert space, one can easily
see that
$$b_i^{\dag}b_j\ket{...\underbrace{0}_j...}=b_i^{\dag}b_j\ket{...\underbrace{1}_i...}=0,$$
\begin{equation}\label{HK}
b_i^{\dag}b_j\ket{...\underbrace{0}_i...\underbrace{1}_j...}=\ket{...\underbrace{1}_i...\underbrace{0}_j...}.
\end{equation}
where, we have used the facts that $b\ket{1}=\ket{0}$,
$b\ket{0}=0$, $b^{\dag}\ket{0}=\ket{1}$ and $b^{\dag}\ket{1}=0$.
Now, let $\ket{k}$ denotes the vector state which its all
components are $0$ except for $k$, i.e.,
$\ket{k}=b_k^{\dag}\ket{{\mathbf{0}}}=\ket{0...0\underbrace{1}_k0...0}$.
Then, it can be easily seen that
\begin{equation}\label{P}
b_i^{\dag}b_j\ket{k}=\delta_{jk}\ket{i} \;\;\ \rightarrow \;\;\
b_i^{\dag}b_j=E_{ij},
\end{equation}
where, $E_{ij}$ is an $n\times n$ matrix all of whose elements are
zero except the $(i,j)$ element which is unity, i.e.,
$(E_{ij})_{kl}=\delta_{ik}\delta_{jl}$. Then, one can easily
deduce that
\begin{equation}\label{P'}
\sum_{_{i{\sim}_k j}}b_i^{\dag}b_j=\sum_{_{i{\sim}_k
j}}E_{ij}=A_k,
\end{equation}
where, $i{\sim}_k j$ means $i$ and $j$ are adjacent in the graph
$\Gamma_k$ or equivalently, the coupling strength between $i$ and
$j$ is $J_k$. Then, by using (\ref{P'}), the hamiltonian in
(\ref{eqH}) can be written in terms of the adjacency matrices
$A_k$, $k=0,1,...,d$ as follows
\begin{equation}\label{HA}
H=\sum_{k=0}^dJ_k\sum_{_{i{\sim}_k j}}E_{ij}=\sum_{k=0}^dJ_kA_k.
\end{equation}
Assume that we are able to obtain the spectrum of $A_k$ so that we
can write $A_k=\sum_{l}\lambda^{(k)}_lE_l$ ($E_l$ is the
projection operator to the subspace associated with the eigenvalue
$\lambda^{(k)}_l$). Then, we have
$$ f_{j1}(t_0)=
\langle {\mathbf{0}}|b_je^{-iHt_0}b^{\dag}_1|
{\mathbf{0}}\rangle=\sum_l\langle {\mathbf{0}}|b_je^{-it_0\sum_k
J_k\lambda^{(k)}_l}E_lb^{\dag}_1|
{\mathbf{0}}\rangle=\sum_le^{-it_0\tilde{J}_l}\langle
{\mathbf{0}}|b_jE_lb^{\dag}_1| {\mathbf{0}}\rangle.$$ By comparing
with (\ref{f}), we obtain \be \label{f1} U^{\dag}_{jl}=\langle
{\mathbf{0}}|b_jE_lb^{\dag}_1| {\mathbf{0}}\rangle=\langle
j|E_l|1\rangle.\ee
\subsection{Generalization to the PST of a qudit state}
Assume that we can prepare a $d$-level quantum state (qudit) as
\be \ket{\phi_0}=\alpha_0\ket{{\mathbf{0}}}+\alpha_1
b_1^{\dag}\ket{{\mathbf{0}}}+\alpha_2
(b_1^{\dag})^2\ket{{\mathbf{0}}}+\ldots +\alpha_d
(b_1^{\dag})^d\ket{{\mathbf{0}}}.\ee Now, using the equations
(\ref{eqH2}) and (\ref{eqH3}), one can write
$$e^{-iHt}(b_1^{\dag})^i\ket{{\mathbf{0}}}=e^{-iHt}\sum_{l_1,\ldots,l_i}{\tilde{b}}^{\dag}_{l_1}{\tilde{b}}^{\dag}_{l_2}\ldots {\tilde{b}}^{\dag}_{l_i}\ket{{\mathbf{0}}}=
\sum_{l_1,\ldots,l_i}e^{-it\sum_{_{k=1}}^i{\tilde{J}}_{l_k}}{\tilde{b}}^{\dag}_{l_1}{\tilde{b}}^{\dag}_{l_2}\ldots
{\tilde{b}}^{\dag}_{l_i}\ket{{\mathbf{0}}}=$$
$$\sum_{k_1,\ldots,k_i}(\sum_{l_1}e^{-it{\tilde{J}}_{l_1}}U^{*}_{l_1k_1})(\sum_{l_2}e^{-it{\tilde{J}}_{l_2}}U^{*}_{l_2k_2})\ldots(\sum_{l_i}e^{-it{\tilde{J}}_{l_i}}U^{*}_{l_ik_i})b^{\dag}_{k_1}\ldots b^{\dag}_{k_i}\ket{{\mathbf{0}}}=$$
$$\sum_{k_1,\ldots,k_i}(U^{\dag}{\mathbf{\tilde{J}}})_{k_1}(U^{\dag}{\mathbf{\tilde{J}}})_{k_2}\ldots (U^{\dag}{\mathbf{\tilde{J}}})_{k_i}b^{\dag}_{k_1}\ldots b^{\dag}_{k_i}\ket{{\mathbf{0}}}$$
where, $\mathbf{\tilde{J}}$ has defined by (\ref{J}). Therefore,
the final state of the system is given by $$
\ket{\phi_0(t)}=e^{-iHt}\ket{\phi_0}=\alpha_0\ket{{\mathbf{0}}}+\alpha_1
\sum_{k_1}(U^{\dag}{\mathbf{\tilde{J}}})_{k_1}b_{k_1}^{\dag}\ket{{\mathbf{0}}}+\alpha_2
\sum_{k_1,k_2}(U^{\dag}{\mathbf{\tilde{J}}})_{k_1}(U^{\dag}{\mathbf{\tilde{J}}})_{k_2}b_{k_1}^{\dag}b_{k_2}^{\dag}\ket{{\mathbf{0}}}+\ldots+$$\be\alpha_d
\sum_{k_1,\ldots,k_d}(U^{\dag}{\mathbf{\tilde{J}}})_{k_1}\ldots(U^{\dag}{\mathbf{\tilde{J}}})_{k_d}b_{k_1}^{\dag}\ldots
b_{k_d}^{\dag}\ket{{\mathbf{0}}} \ee Now, in order that PST from
the first site to the $m$-th one be achieved, i.e., we can obtain
the evolved state $\ket{\phi_0(t)}$ as \be
\ket{\phi_0(t)}=\alpha_0\ket{{\mathbf{0}}}+\alpha_1
b_m^{\dag}\ket{{\mathbf{0}}}+\alpha_2
(b_m^{\dag})^2\ket{{\mathbf{0}}}+\ldots +\alpha_d
(b_m^{\dag})^d\ket{{\mathbf{0}}},\ee we should have the constraint
\be(U^{\dag}{\mathbf{\tilde{J}}})_{i}=e^{i\theta}\delta_{im}\ee
which is the same condition obtained for the purpose of PST of a
qubit ($d=2$). This indicates that, by choosing suitable coupling
strengths $J_l$, $l=0,1,\ldots, d$, one can transfer a qubit, a
qutrit, and in general a qudit, simultaneously.
\section{Underlying networks of group schemes} Now, we consider some
special graphs which are defined via a finite group $G$. These
graphs have the preference that, the adjacency matrices $A_k$ are
simultaneously diagonalizable and the needed information about the
matrix $U$ can be obtained via the group characters. In order to
define these graphs, first we recall the notion of an association
scheme. For more details about association schemes and their
underlying networks, refer to  \cite{pstass}, \cite{Ass.sch.},
\cite{js} and \cite{jss2}.

Assume that $V$ and $E$ are vertex and edge sets of a regular
graph, respectively. Then, the matrices $A_i$ for $i=0,1,\ldots,
d$ form a commutative association scheme with diameter $d$ if
\begin{equation}\label{ss}
A_iA_j=\sum_{k=0}^{d}p_{ij}^kA_{k},
\end{equation}
$A_0=I$, and the sum of $A_i$ is the all-one matrix $J$. From
(\ref{ss}), it is seen that the adjacency matrices $A_0, A_1,
\ldots, A_d$ form a basis for a commutative algebra \textsf{A}
known as the Bose-Mesner algebra of the association scheme. This
algebra has a second basis $E_0,..., E_d$ (known as primitive
idempotents) so that
\begin{equation}\label{idem}
E_0 = \frac{1}{N}J, \;\;\;\;\;\;\ E_iE_j=\delta_{ij}E_i,
\;\;\;\;\;\;\ \sum_{i=0}^d E_i=I.
\end{equation}
 Let $P$ and $Q$ be the matrices relating the two bases for
$\textsf{A}$:
$$
A_i=\sum_{j=0}^d P_{ij}E_j, \;\;\;\;\ 0\leq i\leq d,
$$
\begin{equation}\label{m2}
E_i=\frac{1}{N}\sum_{j=0}^d Q_{ij}A_j, \;\;\;\;\ 0\leq i\leq d.
\end{equation}
Then, clearly we have
$$A_iE_j=P_{ij}E_j,$$
\begin{equation}\label{pq}
PQ=QP=NI.
\end{equation}
which shows that the $P_{ij}$ is the $j$-th eigenvalue of $A_i$
and that the columns of $E_j$ are the corresponding eigenvectors.
Thus, $m_i=$rank$(E_i)$ is the multiplicity of the eigenvalue
$P_{ij}$ of $A_i$ (provided that $P_{ij}\neq P_{kj}$ for $k \neq
i$).
\subsection{Group association schemes}
Group schemes are particular association schemes for which the
vertex set contains elements of a finite group $G$ and the $i$-th
adjacency matrix $A_i$ is defined as:
$$A_i={\bar{C_i}}:=\sum_{g\in C_i}g,$$
where  $C_0 = \{e\},C_1, ...,C_d$ are the conjugacy classes of $G$
and $g$ is considered in the regular representation of the group.
The corresponding idempotents $E_0, ...,E_d$ are the projection
operators as
\begin{equation}E_{k}=\frac{\chi_{k}(1)}{|G|}\sum_{\alpha\in G}\chi_{k}(\alpha^{-1})\alpha\end{equation}
where, $\chi_k$ is the $k$ th irreducible character of $G$. Thus
eigenvalues of adjacency matrices $A_{k}$ and idempotents $E_{k}$
are given by
\begin{equation}\label{eqx}P_{ik}=\frac{\kappa_i\chi_{k}(\alpha_{i})}{d_{k}},\;\;\;\
Q_{ik}=d_{i}\overline{\chi_{i}(\alpha_{k})}\end{equation}
respectively, where $d_{j}=\chi_{j}(1)$ is the dimension of the
irreducible character $\chi_j$ and $\kappa_k\equiv|C_k|$ is the
$k$th valency of the graph.

By using (\ref{m2}) and (\ref{eqx}), the result (\ref{f1}) is
written as \be\label{u} U^{*}_{lm}=\langle
m|E_l|\phi_0\rangle=\frac{1}{|G|}Q_{lm}=\frac{1}{|G|}d_l{\bar{\chi}}_l(\alpha_m)
\ee Then, the PST condition (\ref{eqH6}) takes the form
 \be
\label{eqH6'}\sum_{l=0}^dU^{*}_{lm}{{\mathbf{\tilde{J}}}}_l=
\frac{1}{|G|}\sum_{l=0}^dd_l{\bar{\chi}}_l(\alpha_m)e^{-it_0{\tilde{J}}_l}=e^{i\theta}.
                       \ee
                       Or equivalently,\be
\label{eqH66}
\frac{1}{|G|}\sum_{l=0}^dd_l{\bar{\chi}}_l(\alpha_m)e^{-i(t_0{\tilde{J}}_l+\theta)}=1.
                       \ee
Using the fact that $\sum_{l=0}^dd^2_l=|G|$, the Eq.(\ref{eqH66})
gives \be \label{eqH66'}
e^{-i(t_0{\tilde{J}}_l+\theta)}=\frac{d_l}{{\bar{\chi}}_l(\alpha_m)}.
                       \ee
It should be noticed that, $\alpha_m\in C_m$ belongs to the center
of the group (and so commutes with all elements of the group);
Then, the well known Schur's lemma in the group representation
theory implies that $\alpha_m$ is represented by $\mu \mathbf{1}$
for some $\mu\in \mathcal{C}$. Assume that $\alpha_m$ has order
$r$ ($r$ is the smallest positive integer for which, we have
$\alpha^r_m=1$). Then, clearly $\mu$ must be the $r$-th root of
unity, i.e., $\mu=e^{2\pi i/r}$ and so $|\mu|=1$. consequently, we
have $|\chi_k(\alpha_m)|=|\mu|d_k=d_k$ and from the fact that
$\chi_k(\alpha_m)$ is real, then we have
$\frac{d_l}{{\bar{\chi}}_l(\alpha_m)}=\pm1=e^{-i\pi n_l}$, $n_l\in
\mathcal{Z}$. Then, the result (\ref{eqH66'}) gives
\be\label{eqH6'} {\tilde{J}}_l=\frac{\pi n_l-\theta}{t_0}, \;\;\
l=0,1,\ldots, d \;\ ;\;\ n_l\in \mathcal{Z}. \ee

From the fact that the adjacency matrices $A_k$ possess $d+1$
distinct eigenvalues (see Eq.(\ref{pq})) given by
$\lambda^{(k)}_l=P_{kl}=\frac{\kappa_k\chi_l(\alpha_k)}{d_l}$ for
$l=0,1, \ldots, d$, we have only $d+1$ distinct ${\tilde{J}}_{l}$
given by
\begin{equation}\label{eqH4'}
{\tilde{J}}_{l}=\sum_{k=0}^{d}J_kP_{kl}.
\end{equation}
The above equation implies that \be\label{eqH6''}
\left(\begin{array}{c}
                      {\tilde{J}}_0 \\
                         {\tilde{J}}_1\\
                         \vdots\\
                         {\tilde{J}}_{d} \\
                       \end{array}\right)=\mathrm{P}^t\left(\begin{array}{c}
                      {J}_0 \\
                         {J}_1\\
                         \vdots\\
                         {J}_{d} \\
                       \end{array}\right). \ee
By using the invertibility of the eigenvalue matrix $\mathrm{P}$
(see Eq. (\ref{pq})), the above equation leads to the following
relation for coupling strengths $J_k$:

\be\label{eqH8}
J_l=\sum_{k=0}^{d}({\mathrm{P}^t}^{-1})_{lk}{\tilde{J}}_k=\frac{1}{|G|}\sum_{k=0}^{d}({\mathrm{Q}^t})_{lk}{\tilde{J}}_k=\frac{1}{|G|t_0}\sum_{k=0}^{d}{d}_k{\bar{\chi}}_{k}(\alpha_l){\tilde{J}}_{k}
,\;\ l=0,1,\ldots, d ;\;\  n\in \mathcal{Z }.\ee where, we have
used the equations (\ref{pq}) and (\ref{eqx}). Now, using the
equation (\ref{eqH6'}) we
 obtain an explicit formula for suitable coupling strengths $J_l$
 as
\be\label{eqH8'} J_l=\frac{1}{|G|t_0}\sum_{k=0}^{d}(\pi
n_k-\theta){d}_k{\bar{\chi}}_{k}(\alpha_l),\;\ l=0,1,\ldots, d
;\;\  n_k\in \mathcal{Z }.\ee
\section{Examples}
\subsection{Cyclic graph $C_{2m}$}
The undirected cyclic graph $C_{2m}$ with $2m$ nodes have the
following adjacency matrices
$$A_0=I,\;\ A_i=S^i+S^{-i}, \;\ i=1,2,\ldots, m-1 \;\ ; \;\ A_m=S^m,$$
where $S$ is the circulant matrix of order $n=2m$, i.e.,
$S^{2m}=I$. Then, it is well known that the finite Fourier
transform $F_n$ with matrix entries
$(F_n)_{kl}=\frac{1}{\sqrt{n}}\omega^{kl}$ ($\omega:=e^{\frac{2\pi
i}{n}}$) diagonalizes the above adjacency matrices,
simultaneously. Therefore, the Bose-Hubbard Hamiltonian
(\ref{eqH}) is given by
$$H=\sum_{i,j=1}^{2m}(J_0I+\sum_{k=1}^{m-1}J_{k}(S^k+S^{-k})+J_mS^m)_{ij}b^{\dag}_ib_j=\sum_{i,j=1}^{2m}(J_0I+\sum_{k=1}^{m-1}J_{k}(\omega^k+\omega^{-k})+J_m\omega^m)_{ii}{\tilde{b}}^{\dag}_i{\tilde{b}}_i=$$
$$\sum_{i=0}^{2m-1}(J_0+2\sum_{k=1}^{m-1}J_{k}\cos\frac{\pi ki}{m}+J_m(-1)^i){\tilde{n}}_i=\sum_{i=0}^{2m-1}{\tilde{J}}_i{\tilde{n}}_i.$$

$$\ket{\phi_0}=\alpha{\ket{\mathbf{0}}}+\beta b^{\dag}_1\ket{\mathbf{0}}$$
$$\ket{\phi_0(t_0)}=e^{-it_0H}\ket{\phi_0}=\alpha{\ket{\mathbf{0}}}+\frac{\beta}{\sqrt{2m}}e^{-it_0H}\sum_{i=0}^{2m-1}{\tilde{b}}^{\dag}_i{\ket{\mathbf{0}}}=\alpha{\ket{\mathbf{0}}}+\frac{\beta}{\sqrt{2m}}\sum_{l}e^{-it_0{\tilde{J}}_l}{\tilde{b}}^{\dag}_l\ket{\mathbf{0}}.$$
According to (\ref{eqH2}), we have
${\tilde{b}}^{\dag}_l=\frac{1}{\sqrt{2m}}\sum_{k}\omega^{-lk}b^{\dag}_k$.
Then, for the purpose of PST, it suffices that the probability
amplitude
$$\langle m|\frac{1}{2m}\sum_{l,k}\omega^{-lk}e^{-it_0{\tilde{J}}_l}|k\rangle=\frac{1}{2m}\sum_{l=0}^{2m-1}\omega^{-lm}e^{-it_0{\tilde{J}}_l}.$$
be an arbitrary phase $e^{i\theta}$. This leads to
$${\tilde{J}}_l=\frac{-\pi l-\theta}{t_0},$$
and so, we obtain
$$J_l=\frac{1}{2mt_0}\{-\theta+2\sum_{k=1}^{m-1}(-\pi k-\theta)\cos\frac{\pi
kl}{m}-(-1)^l(\pi m+\theta)\}.$$
\subsection{Dihedral group $D_{2n}$}
The dihedral group $G = D_{2n }$ is generated by two generators
$a$ and $b$ with the following relations:
\begin{equation}D_{2n}=\langle a,b:a^{n}=1,b^{2}=1,b^{-1}ab=a^{-1}\rangle\end{equation}
We consider the case of even $n=2m$, the case of odd $n$ can be
considered similarly. The Dihedral group $D_{2n}$ with even $n=2m$
has $m+3$ conjugacy class so that $C_m=\{a^m\}$.
$$C_0=\{e\},\;\ C_i=\{a^i,a^{-i}\},\;\ i=1,\ldots, m-1,\;\ C_m=\{a^m\},\;\ C_{m+1}=\{a^{2j}b,0\leq j\leq m-1\},$$$$ C_{m+2}=\{a^{2j+1}b,0\leq j\leq m-1\}$$
Then, the group scheme $D_{2n}$ with $n=2m$ have the following
adjacency matrices
$$A_0=I_2\otimes I_n,\;\ A_i=I_2\otimes (S^i+S^{-i}), \;\ i=1,2,\ldots, m-1 \;\ ; \;\ A_m=I_2\otimes S^m,$$
$$A_{m+1}=\sigma_x\otimes (I_{n}+S^2+\ldots+S^{2(m-1)}),\;\ A_{m+2}=\sigma_x\otimes (S+S^3+\ldots+S^{2m-1}).$$
where $S$ is the circulant matrix of order $2m$, i.e., $S^{2m}=I$.
 The character table of $D_{2n}$ with
$n=2m$ is given by \cite{gordon}
$$\begin{tabular}{|c|c|c|c|c|c|}
  \hline
   $D_{2n}$& $e$ & $a^{m}$ & $a^{r}\;\ (1\leq r\leq m-1)$ & $b$ & $ab$ \\
  \hline
  $\chi_{0}$ & 1 & 1 & 1 & 1 & 1 \\
  $\chi_{1}$ & 1 & 1 & 1 & -1 & -1 \\
  $\chi_{2}$ & 1 & $(-1)^{m}$ & $(-1)^{r}$ & 1& -1 \\
  $\chi_{3}$ & 1 & $(-1)^{m}$ & $(-1)^{r}$ & -1 & 1 \\
  $\psi_{j} \;\ (1\leq j\leq m-1)$ & 2 & $2(-1)^j$ & $2\cos(2\pi jr/n)$ &  0 & 0 \\
  \hline
\end{tabular}$$
Then, by using the result (\ref{eqH8'}), we obtain
$$J_0=-\frac{(m+3)(\theta+\pi m/2)}{2nt_0},$$
$$J_1=\frac{1}{2nt_0}\{-(m+1)\theta- \frac{\pi}{2}[(m-2)(m-3)+6]\},$$
$$J_2=J_3=\frac{1}{2nt_0}\{-\theta((-1)^m+\sum_{r=1}^{m-1}(-1)^r)+\pi(1-\sum_{j=1}^{m-3}(-1)^jj)\}$$
$$J_{l+3}=\frac{2}{nt_0}\{-\theta(1+(-1)^l+\sum_{r=1}^{m-1}\cos \frac{2\pi lr}{n})-\pi
\sum_{r=1}^{m-1}r\cos \frac{2\pi lr}{n}\}\} ,$$ \be l=1,2,\ldots,
m-1.\ee
\subsection{Clifford group}The Clifford algebra
with $n$ generator matrices $\gamma_1,\gamma_2,...,\gamma_n$,
obeys the following relations \cite{3}
\begin{equation}\gamma_i\gamma_j+\gamma_j\gamma_i=2\delta_{ij}I\end{equation}
Thus, the $\gamma$'s have square $1$ and anti-commute. The
Clifford group denoted by $CL(n)$ has $2^{n+1}$ elements as
$$CL(n)=\{\pm1, \pm\gamma_{i_{1}} . . .\gamma_{i_{j}}; \;\  i_ 1 < . . . < i_j , j=1,\ldots, n\},$$
where, $i_r\in \{1,2,\ldots, n\}$. We suppose $n > 2 $ throughout.
It is well known that \cite{3}, the center of $CL (n)$ denoted by
$ Z (CL( n))$, consists of $\{\pm1\}$ if $n$ is even and $ \{
\pm1,\pm\gamma_1 ...\gamma_n\} $ if $n$ is odd. $CL( n)$ has $
2^n$ one-dimensional representations, each real. In each such
representation, $U( -1) =I$; Any irreducible representation with
dimension greater than $1$ has $U(-1) =-I$.

For even $n$, the conjugacy classes are given by
$$C_0=\{1\},\;\ C_1=\{-1\},\;\ C_2=\{\gamma_{1},-\gamma_{1}\},\ldots,\;\ C_j=\{\gamma_{i_1}...\gamma_{i_j},-\gamma_{i_1}...\gamma_{i_j}\},$$
$$C_{2^n+1}=\{\gamma_{1}...\gamma_{n},-\gamma_{1}...\gamma_{1}\},$$
whereas for odd $n$, we have
$$C_0=\{1\},\;\ C_1=\{-1\},\;\ C_2=\{\gamma_{1},-\gamma_{1}\},\ldots,\;\ C_j=\{\gamma_{i_1}...\gamma_{i_j},-\gamma_{i_1}...\gamma_{i_j}\},$$$$C_{2^n+1}=\{\gamma_{1}...\gamma_{n}\},\;\ ,C_{2^n+2}=\{-\gamma_{1}...\gamma_{1}\}.$$
The characters of the $2^n$ one dimensional representations are
given by
$$\chi_k(1)=\chi_k(-1)=1,\;\;\ 0\leq k\leq 2^n-1,$$
$$\chi_{2^n}(\pm \gamma_A)=\pm \delta_{A\emptyset}2^{n/2}\;\ \Rightarrow  \;\ \chi_{2^n}(1)=2^{n/2},\;\ \chi_{2^n}(-1)=-2^{n/2}. $$

Then, by using the result (\ref{eqH8}), we obtain the suitable
coupling strengths as
$$J_0=-\frac{1}{2^{n+1}t_0}\{(2^n+1)\theta +\pi l\},$$
$$J_1=J_2=\ldots=J_{2^{n-1}}=\frac{1}{2^{n+1}t_0}(-\theta +\pi l),$$
$$J_{2^{n-1}+1}=\ldots=J_{2^n-1}=-\frac{1}{2nt_0}(\theta +\pi l),$$
$$J_{2^{n}}=0 , \;\;\ l\in \mathcal{Z}.$$
\subsection{The group $U_{6n}$}
The group $U_{6n}$ is a group of order $6n$ which is defined as
$$U_{6n}=\langle a,b: a^{2n}=b^{3}=1,\;\ a^{-1}ba=b^{-1}\rangle.$$
The $3n$ conjugacy classes of $U_{6n}$ are, for $0\leq r\leq n-1$,
$$\{a^{2r}\},\;\ \{a^{2r}b,a^{2r}b^2\},\;\ \{a^{2r+1}, a^{2r+1}b,a^{2r+1}b^2\}.$$
The character table of $U_{6n}$ is given by \cite{gordon}
$$\hspace{-1cm}\begin{tabular}{|c|c|c|c|c|}
  \hline
   $U_{6n}$ & $a^{2r}$ & $a^{2r}b$ & $a^{2r+1}$\\
  \hline
  $\chi_{j}\;\ (0\leq j\leq 2n-1)$ & $\omega^{2jr}$ & $\omega^{2jr}$ & $\omega^{j(2r+1)}$\\
  $\psi_{k} \;\ (0\leq k\leq n-1)$ & $2\omega^{2kr}$ & $-\omega^{2kr}$& 0 \\
  \hline
\end{tabular}$$
with $\omega:=e^{2\pi i/2n}$. Then by using the character table
and the result (\ref{eqH8}), we have \be\label{eqHu}
J_l=\frac{1}{6nt_0}\{\sum_{k=0}^{2n-1}\chi_k(\alpha_l)(\frac{2\pi
k}{n}-\theta)+2\sum_{k=0}^{n-1}\psi_k(\alpha_l)(\frac{2\pi
k}{n}-\theta)\}, \ee where, we have chosen $\alpha_m=a^2$ and
substituted
$\frac{d_k}{\chi_k(a^2)}=\frac{d_k}{\psi_k(a^2)}=\omega^{-2k}=e^{-\frac{2\pi
ik}{n}}$. By using (\ref{eqHu}), one can obtain
$$J_0=\frac{\pi (5n-3)-6n\theta}{6nt_0},$$
$$J_l=\frac{\pi}{3nt_0}(\sum_{k=0}^{2n-1}k\omega^{-2kl}+4\sum_{k=0}^{n-1}k\omega^{-2kl}),\;\;\ l=1,2,\ldots, n-1,$$
$$J_l=\frac{\pi}{3nt_0}(\sum_{k=0}^{2n-1}k\omega^{-2kl}-2\sum_{k=0}^{n-1}k\omega^{-2kl}),\;\;\ l=n,n+1,\ldots, 2n-1,$$
$$J_l=\frac{\pi}{3nt_0}\sum_{k=0}^{2n-1}k\omega^{-k(2l+1)},\;\;\ l=2n,2n+1,\ldots, 3n-1.$$
For instance, for the case $n=2$ (the group $U_{12}$), the
coupling strengths $J_l$
 are given by
$$J_0=\frac{7\pi-12\theta}{12t_0},\;\ J_1=-\frac{\pi}{2t_0},\;\ J_2=\frac{\pi}{3t_0},\;\ J_3=0,\;\ J_4=J^{*}_5=\frac{\pi(i-1)}{6t_0}.$$
\subsection{The group $V_{8n}$} Let
$n$ be an odd positive integer. The group $V_{8n}$ is a group of
order $8n$ which is defined as
$$V_{8n}=\langle a,b: a^{2n}=b^{4}=1,\;\ ba=a^{-1}b^{-1},\;\ b^{-1}a=a^{-1}b\rangle.$$
The $2n+3$ conjugacy classes of $V_{8n}$ are
$$\{1\}, \;\ \{b^2\},\;\ \{a^{2r+1},a^{-(2r+1)}b^2\}\;\ (0\leq r\leq n-1),$$
$$\{a^{2s},a^{-2s}\},\;\ \{a^{2s}b^2,a^{-2s}b^2\}\;\ (1\leq s\leq \frac{n-1}{2}),$$
$$\{a^jb^k : j\;\ \mbox{even},\;\ k=1 \;\ or \;\ 3\},\;\;\ \mbox{and}\;\;\ \{a^jb^k : j\;\ \mbox{odd},\;\ k=1 \;\ or \;\ 3\}.$$
The character table of $V_{8n}$ is given by \cite{gordon}
$$\hspace{-1cm}\begin{tabular}{|c|c|c|c|c|c|c|c|}
  \hline
   $V_{8n}$ & $e$ & $b^{2}$ & $a^{2r+1}(0\leq r \leq n-1)$ & $a^{2s}(1\leq s \leq (n-1)/2)$ & $a^{2s}b^2$ & $b$&$ab$\\
  \hline
  $\chi_{0}$ & 1 & 1 & 1 & 1 & 1 & 1 & 1\\
  $\chi_{1}$ & 1 & 1 & 1 & 1 & 1 &-1 &-1\\
  $\chi_{2}$ & 1 & 1 &-1 & 1 & 1 & 1 &-1\\
  $\chi_{3}$ & 1 & 1 &-1 & 1 & 1 &-1 & 1\\
  $\psi_{j} \;\ (0\leq j\leq n-1)$ & 2 & -2& $2i\sin(2\pi j(2r+1)/n)$ & $2\cos(4\pi j s/n)$& $-2\cos(4\pi j s/n)$& 0 & 0 \\
  $\phi_{j} \;\ (1\leq j\leq n-1)$ & 2 & 2& $2\cos(2\pi j(2r+1)/n)$ & $2\cos(2\pi j s/n)$& $2\cos(2\pi j s/n)$& 0 & 0 \\
  \hline
\end{tabular}$$
where, $\omega:=e^{2\pi i/2n}$. Again, by using the character
table and the result (\ref{eqH8}), we obtain \be\label{eqHv}
J_l=\frac{1}{4nt_0}\{-\theta[2+\sum_{k=0}^{n-1}\bar{\psi}_k(\alpha_l)+\sum_{k=1}^{n-1}\bar{\phi}_k(\alpha_l)]+\pi\sum_{k=0}^{n-1}\bar{\psi}_k(\alpha_l)\},\;\
\ee where, we have chosen $\alpha_m=b$ and substituted
$\frac{d_k}{\chi_k(b)}=\frac{d_k}{\phi_k(b)}=-\frac{d_k}{\psi_k(b)}=1$.
By using (\ref{eqHv}), one can obtain
$$J_0=\frac{\pi-2\theta}{2t_0},\;\;\ J_1=-\frac{\pi}{2t_0},$$
$$J_l=\frac{1}{2nt_0}(\theta-\pi i\sum_{k=1}^{n-1}\sin \frac{2\pi k(2l-3)}{n}),\;\;\ l=2,\ldots, n+1,$$
$$\hspace{-1.5cm}J_l=\frac{1}{2nt_0}\{-\theta\sum_{k=0}^{n-1}[\cos \frac{4\pi k(l-n-1)}{n}+\cos \frac{2\pi k(l-n-1)}{n}]+\pi\sum_{k=0}^{n-1}\cos \frac{4\pi k(l-n-1)}{n}\},\;\;\ l=n+2,\ldots, \frac{3n+1}{2},$$
$$\hspace{-1.5cm}J_l=\frac{1}{2nt_0}\{\theta[\sum_{k=0}^{n-1}\cos \frac{4\pi k(l-\frac{3n+1}{2})}{n}-\cos \frac{2\pi k(l-\frac{3n+1}{2})}{n}]-\pi\sum_{k=0}^{n-1}\cos \frac{4\pi k(l-\frac{3n+1}{2})}{n}\},\;\;\ l=\frac{3(n+1)}{2},\ldots, 2n,$$
$$J_{2n+1}=J_{2n+2}=0.$$
In the case $n=3$ (the group $V_{24}$), the coupling strengths
$J_l$ are given by
$$J_0=\frac{\pi-2\theta}{2t_0},\;\ J_1=-\frac{\pi}{2t_0},\;\ J_2=J_3=J_4=\frac{\theta}{6t_0},\;\ J_5=J_6=J_7=J_8=0.$$
\section{Conclusion}
Perfect state transfer of a qudit in boson networks was
investigated where, a family of Hamiltonians related to the
Bose-Hubbard model is defined which enable PST of an arbitrary
qudit state. By choosing the underlying networks of finite group
schemes as boson networks (i.e., with each vertex (site) of the
network, a bosonic number operator for the bosons located at that
site, is associated), we showed how to perfectly transfer,
arbitrary qudit states in interacting boson lattices. In fact, by
employing the group theory properties of these networks, an
explicit analytical formula for coupling constants in the
Hamiltonians was given, so that the state of a particular qudit
initially encoded on one site can be perfectly evolved to the
opposite site without any dynamical control. Finally, PST on
underlying networks associated with some finite groups was
considered in details.

\end{document}